\documentclass[12pt,onecolumn,showpacs,amssymb,aps,nofootinbib,floatfix]{revtex4-1}
\usepackage{amssymb}
\usepackage{geometry}
\usepackage[english]{babel}
\usepackage{mathrsfs}
\usepackage{epsfig}
\usepackage{epstopdf}
\usepackage{pgfplots}
\usepackage{amsmath}
\usepackage{float}
\usepackage{appendix}
\usepackage{dsfont}
\usepackage{framed,color}
\definecolor{shadecolor}{rgb}{0.7,0.75,0.71}

\usepackage[utf8]{inputenc}
\usepackage{tikz}
\usetikzlibrary{positioning,arrows}
\usetikzlibrary{decorations.pathmorphing}
\usetikzlibrary{decorations.markings}
\usepackage{cancel}
\usepackage{mathtools}

\newcommand{\be}{\begin{eqnarray}}
\newcommand{\ee}{\end{eqnarray}}
\newcommand{\Qsx}[1]{Q_s(x)}

\newcommand{\redd}[1]{\color{black} #1\color{black}}  

\newcommand{\ave}[1]{\left\langle #1 \right\rangle}
\newcommand{\order}[1]{ \mathcal{O} \left( #1 \right) }

  \newcommand{\lqcd}{\Lambda_{QCD}}

\begin{document} \hbadness=10000
\topmargin -0.8cm\oddsidemargin = -0.7cm\evensidemargin = -0.7cm
\title{Azimuthal instabilities of the Gribov-Levin-Ryskin equation}
\author{Guillermo Gambini, Giorgio Torrieri}
\affiliation{IFGW, State University of Campinas, Campinas, Brazil }
\date{\today}


\begin{abstract}
We introduce the phenomenology of elliptic flow in nuclear collisions, and argue that its scaling across energies, rapidities and system sizes could be suggestive of a QCD-based rather than a hydrodynamical explanation.
As a hypothesis for such an explanation, 
we show that the GLR equation develops unstable modes when the parton distribution function is generalized to depend on azimuthal angle.
This generally means that the structure function acquires an azimuthal dependence.
We argue that this process is a plausible alternative explanation for the origin of elliptic flow, one that naturally respects the scaling experimentally observed.
\end{abstract}
\maketitle


\section{\label{phenointro}A phenomenological introduction}

Relativistic nearly-ideal hydrodynamics \cite{expreview,kodama} famously provides a very good quantitative description of azimuthal correlations in high energy nuclear collisions, usually parametrized by $v_n$ coefficients

\begin{equation}
\label{v2def}
\frac{dN}{dp_T dy d\phi} = \frac{dN}{2\pi dp_T dy}\left[1 + 2 \sum_n v_n(p_T,y) \cos\left(n\left(\phi-\Psi_n \right)  \right) \right],
\end{equation}

where $\Psi_n$ is the reaction plane, determined by geometry.

  The quantitative precision of this description, now extending to many Fourier coefficients, has motivated the consensus that an ``ideal fluid'' has been created in heavy ion collisions, for which hydrodynamics is a good effective theory even at $\sim fm$ scales.

However, several phenomenological puzzles have accumulated challenging hydrodynamics as the origin of $v_n$.

\begin{itemize}
\item The near independence of $v_2$, and particularly $v_3$, on system size as long as one of the systems is a nucleus \cite{cms1,cms2}. This means $pA$ and $dA$ collisions \cite{cmspa,phenixda} give, scaled for geometry, the same azimuthal coefficients as $AA$ collisions\cite{phobos}. Recently, even $pp$ collisions have been reported to give comparable coefficients \cite{ppatlas,ppcms}.

\item The near independence of $v_2(p_T)$ on energy and rapidity  \cite{stars,brahms}, up to very high $p_T$ \cite{cmsv2,v2highpt}.
  Hydrodynamic and transport simulations usually present momentum-integrated $v_2$, but energy scan simulations where $v_2(p_T)$ is calculated usually fail \cite{heinzscaling2,sorensen}, as transverse flow develops differently in each energy and influences elliptic flow in each bin.

\item The photon and dielectron $v_2$ are comparable to the hadron $v_2$. Usual explanations aim to provide an {\em enhancement} of this $v_n$, but never to explain the {\em equality} \cite{photon}.
\end{itemize}

\begin{figure}
\begin{center}
  \epsfig{width=0.7\textwidth,figure=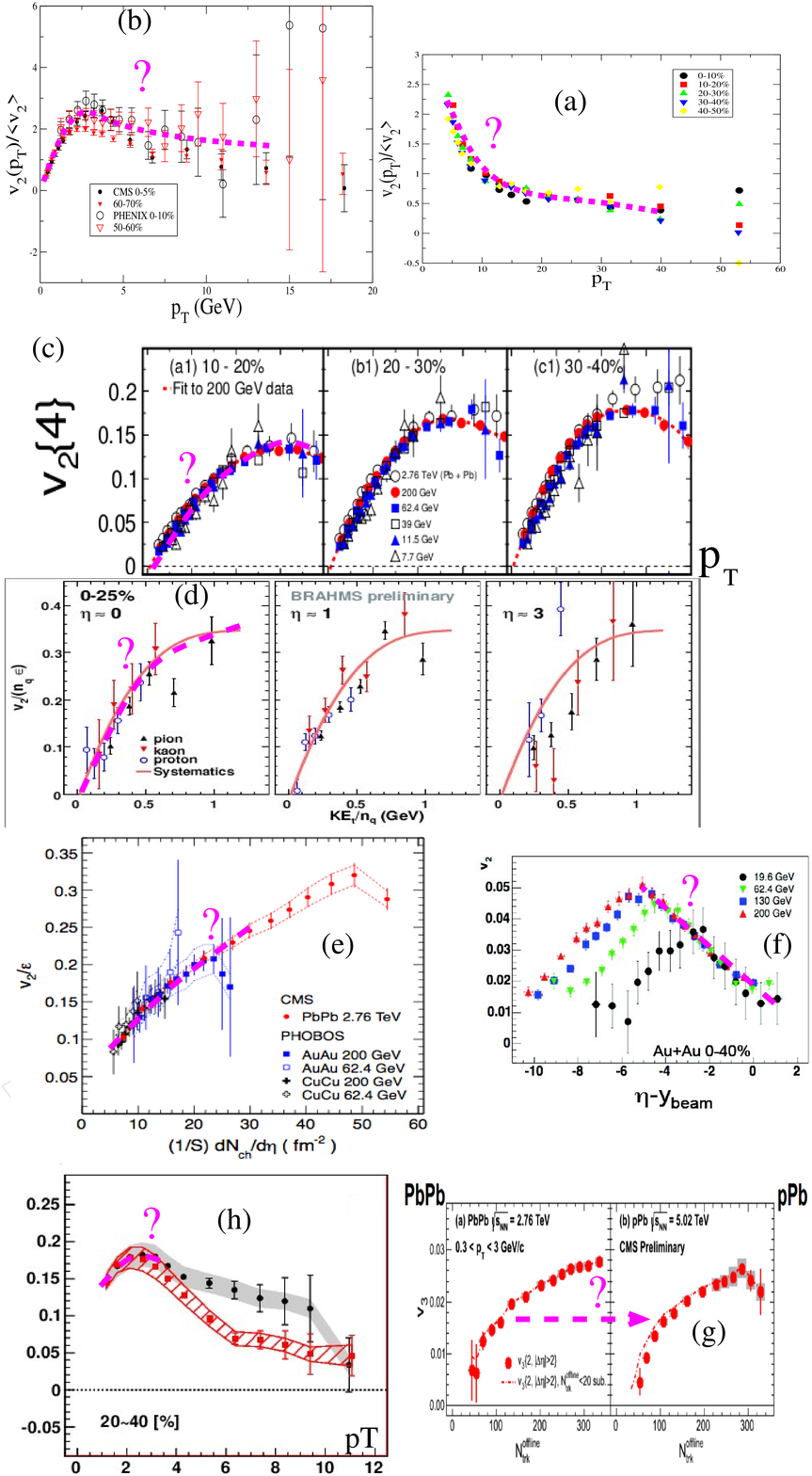}
\end{center}
\caption{
A compilation of the \redd{e}xperimental puzzles of heavy ion collisions described in this section.  Panels (a,b) show the near-same dependence on geometry of, respectively, soft and hard $v_2(p_T)$ \cite{cms1,cms2}.  Panels (c,d) show $v_2(p_T)$ independence on energy and rapidity \cite{stars,brahms}.  (e,f) show that once $p_T$ was integrated, $v_2$ only depends on the transverse entropy density in the same way as $\ave{p_T}$ \cite{cms1,phobos}, while (g,h) show, respectively, the same $v_3$ for pA,dA and AA \cite{cmspa,cmspp,phenixda} and the near-equality of photon and hadron $v_2$ \cite{photon}.}
\label{puzzles}
\end{figure}
Summarizing these \cite{mescaling1,mescaling2,gtridge,v2highpt}, 
\begin{equation}
  \label{scale1}
v_n(p_T) \sim \epsilon_n F(Q \sim p_T) 
\end{equation}
\begin{equation}
  \label{scale2}
v_n \sim \epsilon_n v_n \left( \ave{Q}(\sqrt{s},y,A \right).
\end{equation}

Individually, all these phenomena have been studied, and sometimes even predicted, either by using particular assumptions within hydrodynamic and transport simulations, or by assuming that correlations in small and large systems are of different origin \cite{basar,bozek,raju,galephoton,linnykphoton}. 
However, taken together they represent a compilation of puzzles which are, at the very least, interesting.

In particular, it has long been pointed out, both by heuristic arguments \cite{mescaling1,mescaling2} and explicit simulations \cite{heinzscaling,heinzscaling2}, that the patterns above pose a problem for the hydrodynamic interpretation of $v_2$.

Close to the hydrodynamic limit, one expects $v_2$ to be \redd{:}
\begin{itemize}
\item \redd{a}pproximately $\propto \epsilon$, since $v_2 (\epsilon=0 )=0$, and $\epsilon$ is small and dimensionless.
\item \redd{a}pproximately $\propto c_s(T)$, since $v_2 (c_s=0 )=0$, and the dimensionless $c_s$ tracks the equation of state.

\item \redd{m}aximum for ideal hydrodynamics, since the Knudsen number $Kn$, quantifying the ratio of the mean free path to the system size, is small and dimensionless, $v_2 \sim v_2^{ideal} (1-Kn)$.    In turn, the Knudsen number is related to the viscosity over entropy density $\eta/s$ as well as the system size $R$, i.e. $Kn \sim \eta/(sTR)$.
\item $v_2^{ideal}$ is a highly non-linear function of the lifetime $\tau_{life}$, $v_2^{ideal} \sim v_2(\tau_{life}/\tau_0 \rightarrow \infty) \times f(\tau_{life}/\tau_0)$, which can be numerically shown to be monotonically saturating, $\sim f(\ave{p_T}) \tanh(...)$ in a Cooper-Frye freezeout. $\tau_{life}$ is in turn related to the freezeout temperature and energy density $T_f,e_f$.
\item For $\mu_B \ll m_p$ and isothermal freeze-out, $\tau_{life}/\tau_0 \sim (e_{0}/e_f)^{4 \alpha}$, with $\left. \frac{1}{3} \right|_{bjorken} < \alpha < \left. 1 \right|_{hubble}$ depending on how ``three dimensional'' is the flow, this relation becomes more complicated, but qualitatively similar for systems at high chemical potential.
  \end{itemize}
In summary, elliptic flow in the hydrodynamic limit should scale as

\begin{equation}
\frac{v_2}{\epsilon} \sim c_s  f \left( \frac{1}{T_f^3 \tau_0 R^2}\frac{dN}{dy}  \right) \left( 1 - \order{1} \frac{\eta}{s}  \frac{1}{ T R} \right).  
\end{equation}

It is clear that only $\order{Kn}$ terms mix intensive quantities such as the energy density $e$ with extensive ones such as the size $R$.    $\order{Kn^0}$ ``ideal'' terms, except for the initial time $\tau_0$, depend purely on intensive quantities, giving rise to scaling between systems of different sizes.

Under this scenario, the scaling of $v_2(p_T)$ with energy is puzzling enough, since $v_2$ should increase with energy in {\em all} $p_T$ bins, \redd{as }  transverse flow, increasing for larger and hotter systems in proportion to their longer lifetime, enhances the effect of a given anisotropy at all $p_T$ bins (this is demonstrated in calculations such as \cite{heinzscaling2,sorensen}, where $v_2(p_T)$ rather than integrated $v_2$ is calculated for different energies). 
What is also remarkable is that these scalings only work when $v_n$ summed over all particle species is considered.  $v_n^i(p_T)$ for $i=\pi,K,p,\overline{p}$, etc. certainly does not scale with energy \cite{stars}.   This presents a further puzzle when one considers that the total $v_2$ is a sum of species $v_2$ weighted by relative abundance

\begin{equation}
\label{masscaling}
v_2(p_T) = \frac{\sum_i n_i(p_T) v_2^i(p_T)}{\sum_i n_i(p_T)} =  \frac{v_\pi^2(p_T) n_\pi(p_T) \sum_i \frac{n_i(p_T)}{n_\pi(p_T)}\frac{v_2^i(p_T)}{v_2^\pi(p_T)}}{\sum_i n_i(p_T)}. 
\end{equation}
Now, how flow anisotropy is shared between different species is, if one believes the Cooper-Frye formula \cite{cooperfrye}, independent of the relative abundance of these particles, i.e. it just depends on the masses $m_i$ of the particles.
This means that, while $v_2^{total}$ depends on $\mu_B$ via the equation of state and the lifetime, the dependence on $\mu_B$ of   $v_2^{i}/v_2^{j}$ is rather weak.
\redd{However}, the relative abundance of particles, $n_i/n_j$, strongly and directly depends on $\mu_B$.  This means that, in the upper-right sum of Eq. \ref{masscaling} the first coefficient is very weakly correlated with the second.
Hence, the overall $v_2(p_T)$ scaling  requires a cancellation of quantities that depend on particle abundances and masses in rather different ways. This is  unnatural, yet this is the scaling we see in experimental data \cite{greco}.

The scaling with system size, from $pp$ to $pA$ and $dA$ systems is truly strange, because one naturally expects the opposite. Transverse geometry in $pp$ and $pA$ collisions is very similar, whereas it is different in $AA$ collisions at the same $dN/dy$ (since these are more dilute but more spread out).  Hence, it is naturally expected that similar $dN/dy$ also has a similar $v_{2,3}$ between $pp$ and $pA$, but different (larger) ones in $AA$ \cite{gtridge}.  Instead, all three systems are comparable, though there is room for $v_{2,3}$ suppression in $pp$.
One could, of course, believe that $v_n$ in peripheral AA, just like pA and pp, is non-hydrodynamic in origin while central $AA$ is hydrodynamic, but then comparison of lower energy central $AA$ with $dA$ and $pA$ collisions at top RHIC/LHC energies, and the observed scaling with $dN/dy$ alone \cite{phenixda}, should give this interpretation pause (a pA, dA energy scan at $RHIC$ would test this conclusion much more tightly than what is currently available).

At higher $p_T$, the hydrodynamic regime is thought to be substituted by the tomographic regime \redd{where } $v_n$ is still generated in the same sign as hydrodynamics.   The scaling variables\redd{, } however, should be very different.  While the hydrodynamic $v_n$ is generated by gradients and suppressed by viscosities, tomographic $v_n$ is generated via integrated energy losses, which in turn depend on a combination of the traversed path and the local density.   The exact combination is model dependent, but can be phenomenologically parametrized into an ``ABC model''

\begin{equation}
-\frac{dE}{d\tau} = f(T,p_T,\tau) \simeq \kappa p^{a} T^{b} \tau^c + \order{\frac{T}{p_T},\frac{1}{T\tau}},
\label{abceq}
\end{equation} 

where $a,b,c$ approximate the dependence of the rate of energy loss on momentum $p$, temperature $T$\redd{, } and proper time $\tau$.  In a collisional dominated parton cascade $a=1,c=0$, in a radiative dilute plasma (``Bethe-Heitler regime'') $a=1/3, c=0$, in a dense plasma (LPM regime) $a=1,c=1$, while in a ``falling string'' AdS/CFT scenario $a=1/3,c>2$ \cite{betz}.  

Hence, one expects the hydrodynamic regime and the tomographic regime to scale very differently with temperature and system size.  The evidence we have\redd{, } however, is that this scaling is remarkably similar across both energies and system sizes.

\redd{I}f this evidence is confirmed and extended to $pA$ collisions, particularly in regard to the critical $p_T$ when $v_n$ turns off \cite{gtridge}, it would imply that low momentum and high momentum $v_n$'s have the same scaling variables, something which fundamentally flies in the face of the ``standard model'' of heavy ion collisions.

All of these puzzles are notable {\em not} because physics seems to be more complicated than our models but, on the contrary, because physics seems remarkably simple across orders of magnitude in energy, system size, and rapidity.
This, and the complication of the current ``standard model'' of heavy ion collisions, makes it desirable to at least attempt to develop a different paradigm of generating $v_n$, one where the scalings described in this section come naturally.  It is the purpose of this work to do so.

We start \cite{greco} by noting the fact that structure functions $f(x,Q^2)$ and fragmentation functions $D_{q \rightarrow i}(z,Q^2) $ naturally follow the scaling suggested by both the overlap of $v_2(p_T)$ and its breakdown by particle species, since both of \redd{them } depend weakly on momentum exchange $Q^2$ but strongly on the rescaled variables $x=p_z^{parton}/E,z=p^{hadron}/p^{parton}$ \cite{qcdcoll}.  In structure functions, $x$ is absorbed into the longitudinal component of momentum, with $Q^2 \sim p_T^2$.  In fragmentation functions, $z \sim p_T$ but unitarity protects the effect of \redd{fragmentation } on all hadrons, $\sum_i\int z dz D_{q \rightarrow i}(z,Q^2)=1$. Together, \redd{they } lead to a particle species dependent $v_2^i(p_T)$ with $\sqrt{s}$, but a much weaker dependence of total $v_2(p_T)$.   

Hence, if we could just assume that either the structure function acquired an azimuthal dependence having the usual Bjorken scaling (strong and fundamentally non-perturbative dependence on $x$, and logarithmically suppressed dependence on $Q^2$), and this azimuthal dependence generated most of $v_2$, all issues described in this introduction would be naturally resolved.    The equality of photon and hadron $v_n$ is due to the fact that $v_n$ is really an initial state effect.   The scalings with $\sqrt{s}$, system size\redd{, } and rapidity would come from the usual scalings of non-perturbative QCD operators when probed at scales $\gg \Lambda_{QCD}$.    Both soft and hard $v_n$ would be regulated by the same physical process.
\redd{I}f the EMC effect has something to do with the observed azimuthal dependence, the hierarchy with system size is also naturally explained.
\redd{Last but not least, } the interplay of particle species would be a consequence of unitarity at fragmentation (a parton has to fragment to {\em something}).

Such a ``simple'' suggestion of course is superficially extremely implausible: QCD has azimuthal symmetry, and the parton structure and fragmentation functions
are based on factorization at high energy scales.   Thus, they are ``universal'' and should not depend on relative angles, even if the target and/or projectile are spatially extended.   Yet, ways compatible with $QCD$ to incorporate azimuthal dependence of structure functions do exist.  We certainly know the Sivers effect generates just such a solution \cite{sivers}.  Such a model for $v_2$ is\redd{, } however\redd{,} most likely untenable because the angle $\Psi$ would have no relation to geometry (it would be a reflection of the relative spin orientation between nuclei)\redd{, } and because it would assign a privileged role for $v_2$ w.r.t. other coefficients.

There are semiclassical QCD models, valid at high occupancy number, where different orientations of semiclassical ``color antennae'' generate azimuthal correlations \cite{kovner,raju,pajares,antennae}, though such mechanisms become negligible in the limit of a large number of randomly-pointing antennae, i.e. in AA systems. \redd{As } mechanisms like this are possible for smaller systems, this begs the question of why pA and AA azimuthal correlations look so similar if the underlying processes are different.

\begin{figure}[!h]
\begin{center}
  \epsfig{width=0.8\textwidth,figure=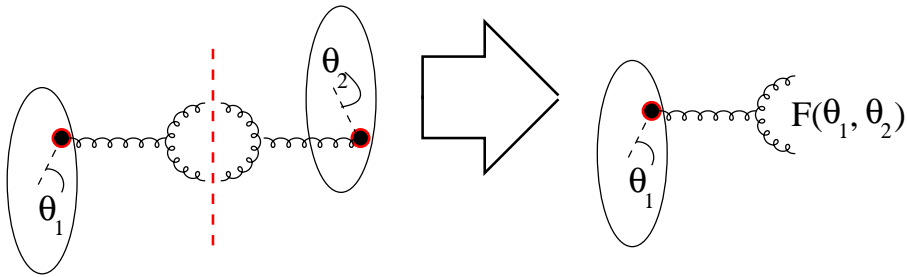}
\caption{Higher twist diagrams leading to an azimuthal dependence of the structure function at higher $x$  \label{impact}}
\end{center}
\end{figure}

More generally \cite{wang}, since the distributions functions depend on the probe used to measure them beyond tree level (``run''), a dipole colliding off-center with an extended object will measure azimuthally asymmetric parton distribution functions to \redd{one-loop precision } (Fig. \ref{impact}).

This point is somewhat subtle, since PDFs are an ``initial state'' property of the hadrons wavefunction, which intuitively should not know about the subsequent evolution of the system. However, within quantum mechanics, the result of the observation in general depends on the observable, even if the observable is an ``initial state''.  In field theory\redd{, } this dependence extends to the scale of the measurement: parton distribution functions depend on the scale $Q^2$ at which they are probed, even though this scale is obviously external to the hadron.  If both the probe and the target are extended, there are two scales, both vector-like: the momentum transfer $Q^\mu$ and the impact parameter $\vec{b}$.  In this regime, Lorentz symmetry will not constrain the dynamics to depend on $Q^2$, for $Q.\vec{b}$ terms, sensitive to nuclear gradients, are allowed.  This \redd{has been } found in explicit calculation\redd{s}, e.g. in \cite{wang}.

\begin{figure}[!h]
\begin{center}
  \epsfig{width=0.5\textwidth,figure=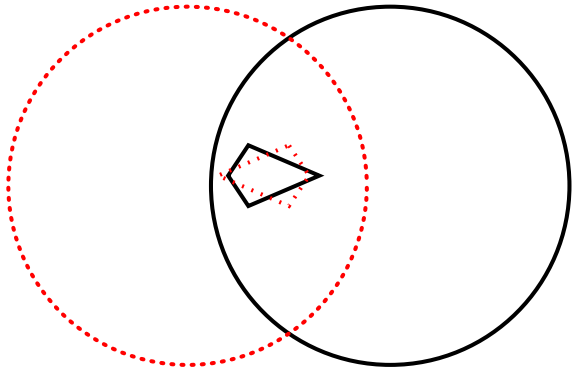}
\caption{ The geometrical picture of how the higher twist contribution generates $v_2$, with the black solid line representing the target and the red line the projectile.  The little polygons sketch the azimuthal phase dependence of the projectile and target structure functions in momentum space}
\label{geometry}
\end{center}
\end{figure}

Assuming nuclear symmetry (projectile=target), each gluon of the projectile hitting the target at a certain impact parameter ``knows'' that the target structure function is slightly ($\order{\alpha_s^2}$) asymmetric w.r.t. the direction\redd{s } up and down of the target nucleus.  This asymmetry must go as $\order{\alpha_s^2} \redd{\hat{r}.\hat{b}} \partial_r f(x,Q)$, where $\hat{r}$ is the unit vector target radial direction, $\hat{b}$ is the impact in the impact parameter direction, and $\partial_r f(x,Q)$ is the radial gradient in the gluon density.
By symmetry, the same effect will happen in the projectile's frame of reference w.r.t. the target, assumed here to be the same kind of nucleus as the projectile.
Any interaction between projectile and target will therefore a\redd{c}quire a  $(f+ \Delta f)(f-\Delta f)$
 dependence (Fig. \ref{geometry}).  It is clear from \redd{this } figure that \redd{such } dependence mirrors exactly the initial gradients used to generate $v_2$ by flow, due to the $\Delta f^2$ term.

Everything said above is easily generalized from an idealized spherically symmetric nucleus to a realistic nucleus with nuclear fluctuations.  In \redd{that } case, the polygons in Fig. \ref{geometry} will point in fluctuating directions, generating, in $\Delta f$, odd harmonics in precisely the same direction as the density gradients in hydrodynamics.   Within hydrodynamics, these density gradients will give rise to pressure gradients and; hence, to flow.   In these processes\redd{; } however, one interaction is enough to produce an anisotropy of final-state particles.

Since the impact parameter $b \sim \order{1-10}fm$, these corrections are expected to be tiny$\sim \alpha_s^2/(bQ)^n$; hence, provided evolution is linear, their contribution to the observed $v_n$ \redd{are smaller } by at least an order of magnitude.  

\begin{figure}[h]
\begin{center}
  \epsfig{width=0.8\textwidth,figure=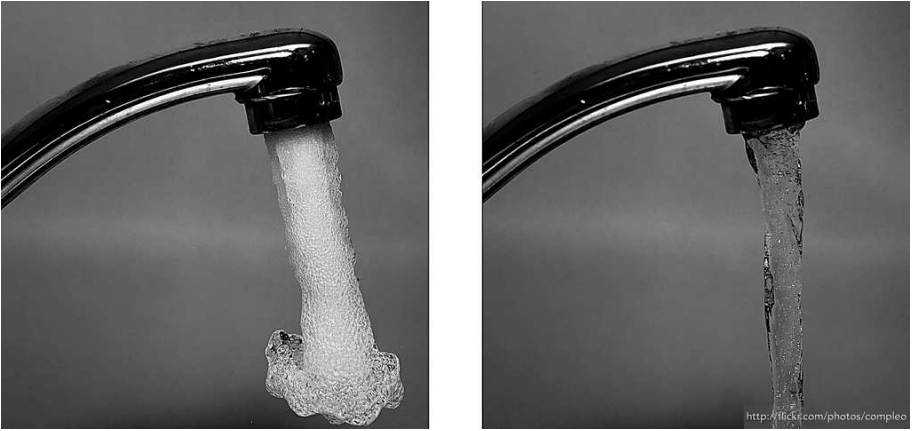}
\caption{\label{faucet}The spontaneous breaking of azimuthal symmetry by turbulent flow of water in a faucet}
\end{center}
\end{figure}

   In this work; however, we investigate \redd{whether } this initial azimuthal symmetry, present at {\em moderate} $x_{projectile}, x_{target}$, can be amplified to lower $x_{projectile}, x_{target}$ (i.e. towards midrapidity)  by the same processes that are thought to lie at the core of saturation physics.  If it \redd{can}, the above anisotropies are indeed tiny at high rapidities, but could be amplified at mid-rapidity, in the same way as cylindrical boundary conditions in a flow of water from a faucet (Fig. \ref{faucet}) lead to an azimuthally asymmetric turbulent flow when density and flow velocity are high enough.   Our analogy here is that the flow is the nucleon's partons flying at the probe at high speed, and the non-linear evolution equation is the one developing turbulence.   

   The resulting azimuthal anisotropy is then the required order of magnitude larger than the higher twist seed, but still ultimately depends on nuclear gradient (as in hydrodynamics and experimental data) and still obeys the scaling laws evident from pQCD (which would explain the experimental scalings seen in the previous section).   In particular, since, kinematically \cite{kovbook}
   \[\  x_{projectile}+ x_{target} \sim \frac{p_T}{\sqrt{s}}  \cosh(y)  \]
an unstable mode in $x$ would grow approximately linearly in $y-y_{beam}$ (where $x_{projectile}\sim 1/3$) up until a maximum depending on $\ln(\sqrt{s})$, reproducing, provided the ``Lyapunov exponent'' is approximately constant (which in an RG evolution it should be), the limiting fragmentation of $v_n$ observed in \redd{experiment} \cite{phobos}. 

In the next few sections, we will give an outline of how this could work.



\section{the GLR equation}
\subsection{Motivation}
The GLR equation \cite{glr2,kovbook} is the simplest non-linear correction to gluon evolution at low $x$. Such non-linear corrections are motivated by the breakdown of unitarity at asymptotically high energy due to the divergence of the number of gluons in linear evolution.

Going beyond this non-linear correction is a focus of extensive and active research.   The most direct consequence of non-linear evolution is the appearance of a Bjorken $x$-specific ``saturation scale'' $Q_s(x) \sim x^{-\lambda}$ \cite{glr1}, where the linear term and the non-linear correction match.
The GLR equation has been interpreted, \redd{in the past, } as evidence that a hierarchy of scales in $\ln(1/x)$ between valence color charges (of small characteristic transverse side and high rapidity) and ``wee gluons'' (of low $x$ and characteristic size $\sim Q_s \gg \lqcd$) require a renormalization group approach.  Such an approach was developed via the JIMWLK and BK equations, and is now firmly established as the ``Color Glass Condensate'' effective theory \cite{cgc}.

However,  the symmetries of the UV theory can in principle be broken in the IR, via the vacuum (``symmetry breaking'') or quantum corrections (``anomalies'').  In these cases, effective theories from the symmetric fixed point will not be realized.
Thus, it is not clear that the action in this limit has all the symmetries of QCD.   In this work, we shall examine, starting from the GLR equation, the possibility of azimuthal symmetry being broken\footnote{To be more exact,  the direction of spontane\redd{o}us symmetry breaking is usually unconnected to initial conditions while here the effective lagrangian will amplify existing nuclear gradient anisotropies.  However, if the \redd{r}enormalization group language is to be used, the effect studied here is a symm\redd{e}try breaking  } by nonlinear dynamics. 

We note that this setup is very different from the idea \redd{of } azimuthally asymmetric CGC fields, proposed earlier in \cite{nara} and developed in \cite{raju}, as well as azimuthally asymmetric solutions of the BK equation \cite{bkb1,bkb2,bkb3}.  In this case, one solves an azimuthally symmetric lagrangian with asymmetric boundary conditions.  We propose that the lagrangian itself dynamically aquires a preferred direction, something that vanishes in the UV.     While both GLR and JIMWLK are renormalization group equations, the first ultimately connects hard and soft (IR and UV) degrees of freedom (whose RG running is implied by QCD renormalizability), while the second connects central rapidity and large rapidity processes, intrinsically assuming that they develop an RG flow (a reasonable assumption, but one which a spontaneous breaking of azimuthal symmetry in the mean field would invalidate).

As an illustration of the above paragraph, an azimuthally dependent structure function is part of the wavefunction of {\em every} parton in the nucleon.   As such, it should show up in every cumulant of $v_n$, just like experimental data seems to be \cite{cmspp}.  Anisotropies described in CGC field-based models, on the other hand, typically decrease with the number of cumulants \cite{cumreview,larryskokov,dusling,raju}.


\subsection{The analysis}
We shall use the GLR equation, as written in \cite{glr1,glr2}, but allow for \redd{$\Xi$ } to depend on azimuthal angles $\phi$ w.r.t. reaction plane in the $k$ direction.   Using $\theta$ as the angle between $l$ and $k$ we have

\[
\frac{\partial \, \Xi (x,k,\phi)}{\partial \ln (1/x)} \, =   - \frac{\alpha_s^2 \, \pi}{S_\perp} \, \left[ \Xi
(x, k,\phi) \right]^2
+\frac{\alpha_s \, N_c}{\pi^2} \int \! \frac{l d
l d\theta}{k^2+l^2 - 2 l k\cos(\theta)}
 \bigg[ \Xi
(x,l,\theta+\phi)
\]

\begin{equation}
\label{glr}
+ \frac{ k^2 \, }{2 l^2 + k^2 -2 k l\cos(\theta)}\Xi
(x, k,\phi) \bigg].
\end{equation}

In the high $Q^2$ limit, this equation can be put in a purely differential "Mueller-Qiu" form (\cite{kovbook} 3.4), with 
\[\   
xG(x,Q^2) = 2\pi \int^{Q^2} \Xi(x,k) d k^2 \Rightarrow xG(x,Q,\phi) = \int^{Q^2} \Xi(x,k,\phi) k d k .
\]

\redd{I}n cylindrical coordinates we have
\begin{eqnarray}\label{mq}
  \frac{\partial^2 \, xG (x, Q^2) }{\partial \ln (1/x)\, \partial \ln
  (Q^2 / \Lambda^2) } \, = \, \frac{\alpha_s \, N_c}{\pi} \, xG (x, Q^2) -
  \frac{\alpha_s^2 \, \pi}{S_\perp} \, \frac{1}{Q^2} \, [xG (x, Q^2)]^2.
\end{eqnarray}

Assuming azimuthal symmetry, everything depends on $Q$ rather than $Q_x,Q_y$, or equivalently in Eq. \ref{glr} $G(x,k,\theta)=G(x,k,\theta+\phi)$, and the angle can be integrated out.   
The most trivial modification one can make is to relax this approximation, so the integral over $\theta$ in Eq. \ref{glr} is not trivial but one still goes into the nearly collinear high $Q^2$ limit.   It is easy to see that this modifies Eq. \ref{mq} by a relatively simple substitution

\[\ \frac{\partial}{\partial Q} \rightarrow \frac{\partial}{\partial Q} + \frac{1}{Q} \frac{\partial}{\partial \theta},  \]

and, up to corrections of the same order that we shall study later in this work, the GLR-MQ equation becomes

\begin{eqnarray}\label{mqphi}
\frac{1}{\Lambda^2} \left( \frac{\partial}{\partial Q} + \frac{1}{Q}\frac{\partial}{\partial \theta}   \right) \frac{\partial \, x Q^2 G (x, Q^2) }{\partial \ln (1/x)}  \, = \, \frac{\alpha_s \, N_c}{\pi} \, xG (x, Q^2) -
  \frac{\alpha_s^2 \, \pi}{S_\perp} \, \frac{1}{Q^2} \, [xG (x, Q^2)]^2,
\end{eqnarray}

We intend to perturb the solutions of (\ref{mqphi}), i.e. $G_0(x,Q^2)$, the following way

\begin{equation}
\label{proposal}
G(x,Q^2,\theta) = G_0(x,Q^2) \left( 1+\sum_n u_n (x,Q^2) cos(n\theta + \beta_n) \right),
\end{equation}

where the background $G_0(x,Q^2)$ is modeled as the saturation scenario, the solution of the azimuthally symmetric non-linear equation, a transcendental function approximately equal to

\begin{equation}
G_0(x,Q^2) \simeq \frac{x^{2\lambda}}{2\alpha_s^4} \left[(1-\tanh(\xi)) + \frac{Q^2_s(x)}{Q^2} (1+\tanh(\xi)) \right],
\end{equation}

with $Q_s(x)=\alpha_s^2 \Lambda_{QCD}x^{-\lambda}$, $\xi=\frac{Q-Q_s(x)}{\zeta}$, and $\zeta \sim \frac{\alpha_s \Lambda_{QCD}}{N_c}$, so that $G_0(x,Q^2)$ has the approximate step function form determined by requiring the two sides of Eq. \ref{glr} to balance

\begin{equation}
\label{qs}
Q_s^2(x) = \frac{\alpha_s \pi^2}{S_\perp N_c} x G_s(x,Q_s(x)) \rightarrow Q_s(x) \sim x^{2 \lambda+1}.
\end{equation}

It is of special interest for us to study parton distribution functions when the probe energy is small compared to the saturation scale; i.e. ${Q}/{Q_s(x)} \ll 1$. 
In this limit we obtain

\[
(2 \lambda +1)  \frac{Q}{2} \frac{\partial u_n(x,Q^2)}{\partial Q} + \frac{Q}{2} x \frac{\partial^2 u_n(x,Q^2)}{\partial Q \partial x} = \bigg[ \frac{\alpha_s N_c}{\pi} +  \frac{ N_c \pi}{C_F S_{\perp}\alpha_s^2} \frac{x^{2\lambda +1}}{Q^2}  \bigg] u_n (x,Q^2)
\]

\begin{equation}
\label{homeq}       
     \hspace{1.2in}+ \delta u_n(x,Q) , 
\end{equation}

with
\[
\delta u_n(x,Q)=   \frac{ N_c \pi}{2C_F S_{\perp}\alpha_s^2} \frac{x^{2\lambda +1}}{Q^2} - \left[  \frac{1}{2} \sum_k^{n-1} u_k(x,Q^2) u_{n-k}(x,Q^2) cos (\beta_n - \beta_{k} - \beta_{n-k} ) \right.
\]

\begin{equation}
\label{uneq}
\left. + \sum_k u_k(x,Q^2) u_{n+k}(x,Q^2) cos(\beta_n + \beta_k -\beta_{n+k}) \right] .
\end{equation}

Assuming $\delta u_n$ is small  (a Taylor expansion can go beyond this approximation), one can study the behavior of linearized instabilities (where the instability interacts with the azimuthally symmetric ``saturating'' component) relatively easily.

We should warn the reader that the equation we are studying is, at best, a rough model.  First of all, it is only a limiting $Q^2$ approximation of an integro-differential equation Eq. \ref{glr}.   This equation was also  derived via fusion of pomeron ladders, which were assumed to also originate from azimuthally symmetric structure functions.   Additional terms, of the \redd{following form,  are also possible}

\begin{equation}
\label{crossladder}
\underbrace{ \int K(\Delta \theta_{12},\Delta x_{12},\Delta Q_{12}) G(x_1,Q_1,\theta_1) G(x_2,Q_2,\theta_2) }_{GLR} \rightarrow \underbrace{F(Q, \phi) \frac{\partial^2}{\partial Q \partial \phi}, \frac{\partial^2}{\partial x \partial \phi} }_{GLR-MQ},
\end{equation}

\redd{where $\Delta \theta_{1,2}=\theta_1-\theta_2-\theta$, $\Delta x_{12}=x_1-x_2-x$, and $\Delta Q_{12}=Q_1-Q_2-Q$. }   In this work, our main aim is to obtain analytical and semi-analytical solutions, so concentrating on a purely differential equation such as Eq. \ref{mq} is justified.  Extensive numerical work will be needed for anything beyond qualitative predictions.


\subsection{\label{secpol}Warmup: a polynomial ansatz}
An ansatz specific for the initial condition set at $x \ll 1$ is the Polynomial one.  Assuming only one Fourier harmonic is non-zero \redd{in this scenario}, we get

\begin{equation}
u_n(x \sim 10^{-1},Q^2) = \epsilon \delta_{n,2} \sum_{k=0}^{\infty} A_k \frac{(Bx^C)^k}{k!}Q^{D-2k}, 
\end{equation}

where the coefficients are easily fixed by algebra in terms of the impact parameter and the parameters of the GLR equation

\begin{equation}
B=\pi^2/C_F S_{\perp}\alpha_s^3, \hspace{0.1in} 
C=2\lambda +1, \hspace{0.1in} 
D=2\alpha_s N_c /(2\lambda +1)\pi, \hspace{0.1in} 
\epsilon = (2R-b)/(2R+b)
\end{equation}

While we shall use the Polynomial ansatz as a comparison, the associated initial condition is,  however, most likely non-physical, since as argued in the previous section, the most likely \redd{form } for anisotropies to be seeded is \redd{to } moderate to high $x$ higher\redd{-}twist processes.   Hence, an ansatz sensitive to initial conditions must be found.


\subsection{\label{secbessel}A Bessel function ansatz with appropriate initial conditions}
One can construct such a solution of eq. \ref{homeq} by \cite{majer} assuming a solution of the form 

\begin{equation}
u(x,Q^2)=x^p f(t)    
\end{equation}

with $t=Bx^C/Q^2  = z^2/2D$. So, we get a Bessel's equation with pure imaginary index (order) $i\nu$ for $f(t)$

\begin{equation}
f^{''}+\frac{1}{z}f^{'}+(1 + \frac{\nu^2}{z^2})f=0,
\end{equation}

for $p=-C$ with $\nu=\sqrt{2D}$. The solution of this equation is inspired by Bessel functions of the first kind \cite{maty}

\begin{equation}
f_{\nu}(z)=A(z) cos (\nu ln z) + B(z) sin (\nu ln z).
\end{equation}

Here

\[
A(z)=\sum_{n=0}^{\infty} A_{2n} \left( \frac{z}{2} \right)^{2n}, \hspace{0.5in} B(z)=\sum_{n=0}^{\infty} B_{2n} \left( \frac{z}{2} \right)^{2n},
\]

with

\[
A_{2n}=- \frac{nA_{2n-2} - \nu B_{2n-2}}{n(n^2+\nu^2)}, \hspace{0.5in} B_{2n}=- \frac{\nu A_{2n-2} +n B_{2n-2}}{n(n^2+\nu^2)}
\]

for $n\geq 1$ and $A_0$, $B_0$ known. Finally, our solution reads

\begin{equation}
u(x,Q^2)=x^{-(2\lambda +1)}  f_{\nu} \left( z \right)
\end{equation}

with

\begin{equation}
z=\sqrt{ \left( \frac{4\pi N_c}{(2\lambda +1)C_F  \alpha_s^2} \right) \frac{1}{S_{\perp}} \frac{x^{2\lambda +1}}{Q^2}}, \hspace{0.5in} \nu = \left( \frac{4\alpha_s N_c}{\pi(2\lambda +1)} \right)^{1/2}
\end{equation}


\subsection{Results}
The Polynomial and Bessel function ansatzes are incompatible, for one goes to zero while the other diverges at $x \rightarrow 0$.  We therefore think of them as independent solutions, triggered by different initial conditions, and examine the behavior of each case in detail.

\begin{figure}
\begin{center}
  \epsfig{width=0.6\textwidth,figure=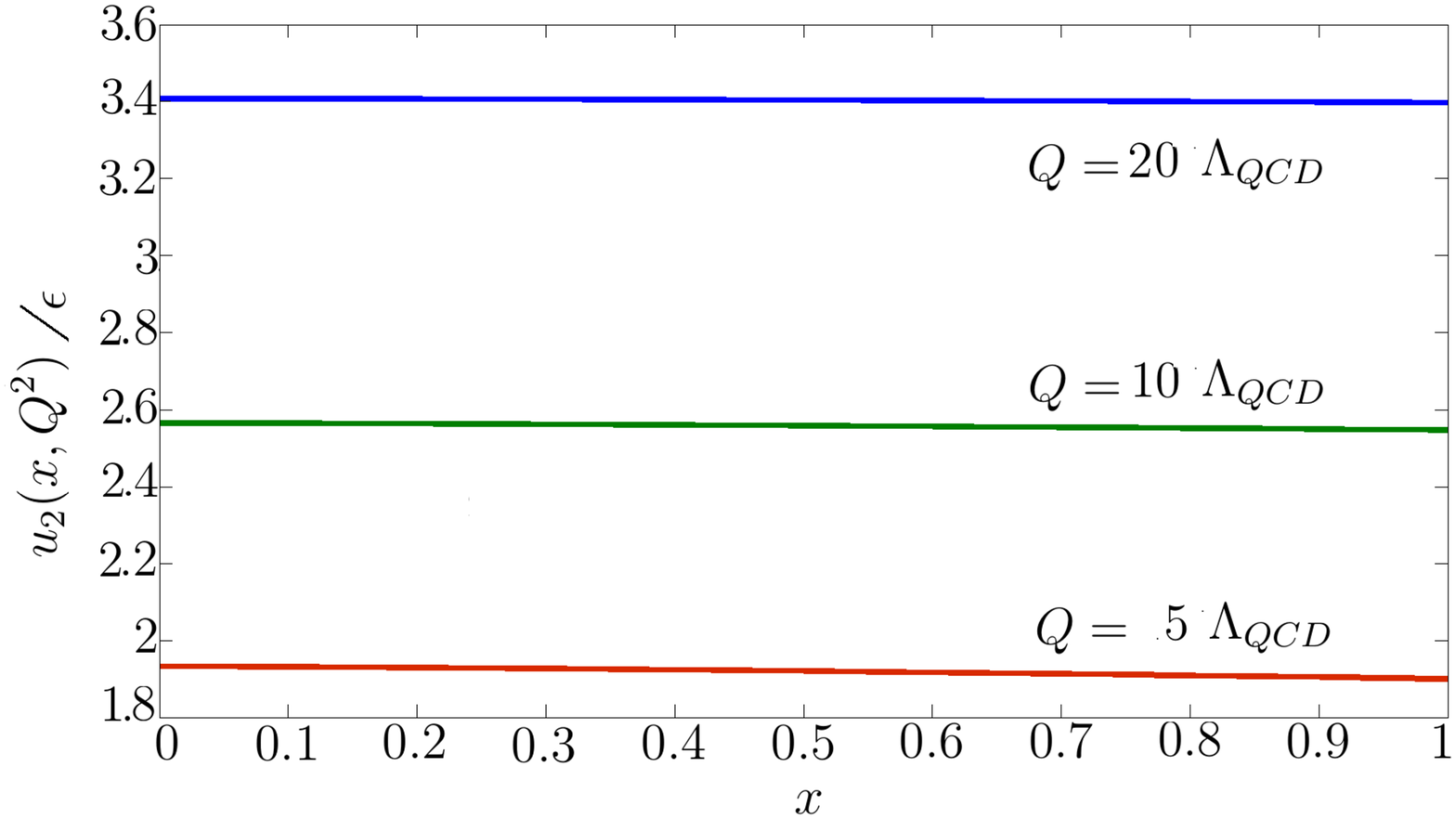}
 \caption{$u_2/ \epsilon $ vs. $x$ according to the Polynomial initial condition.}
\end{center}
\end{figure}

\begin{figure}
\begin{center}
  \epsfig{width=0.6\textwidth,figure=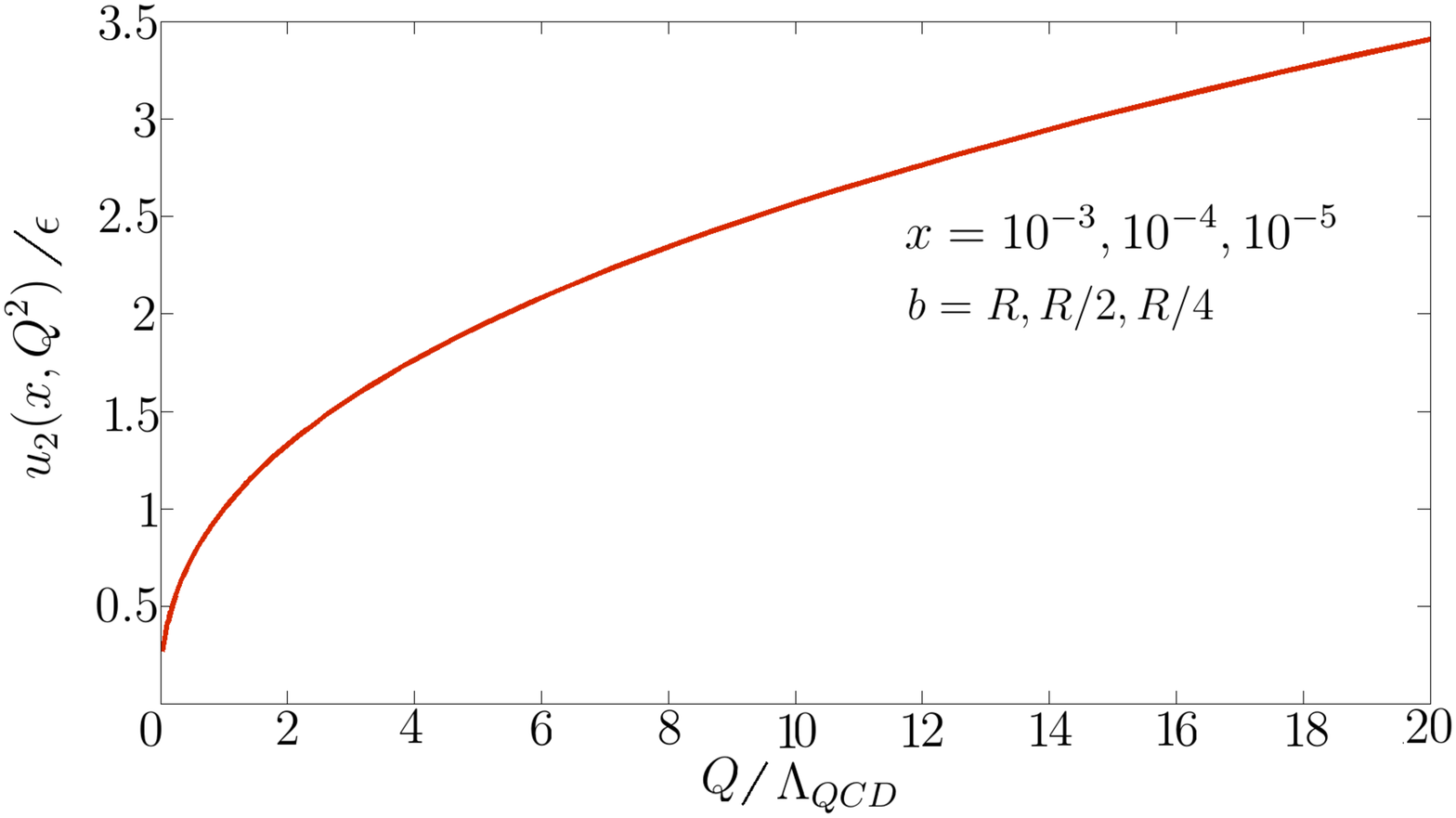}
\caption{$u_2/ \epsilon $ vs. $Q$. Notice that the graphics overlap when going to smaller x or when varying the impact parameter, according to the Polynomial initial condition. }
\end{center}
\end{figure}

We note in that the polynomial ansatz  $u_n(Q)$ is practically constant in $x$, reflecting most of the scalings we examined in the last section.

\begin{figure}[h]
\begin{center}
  \epsfig{width=0.8\textwidth,figure=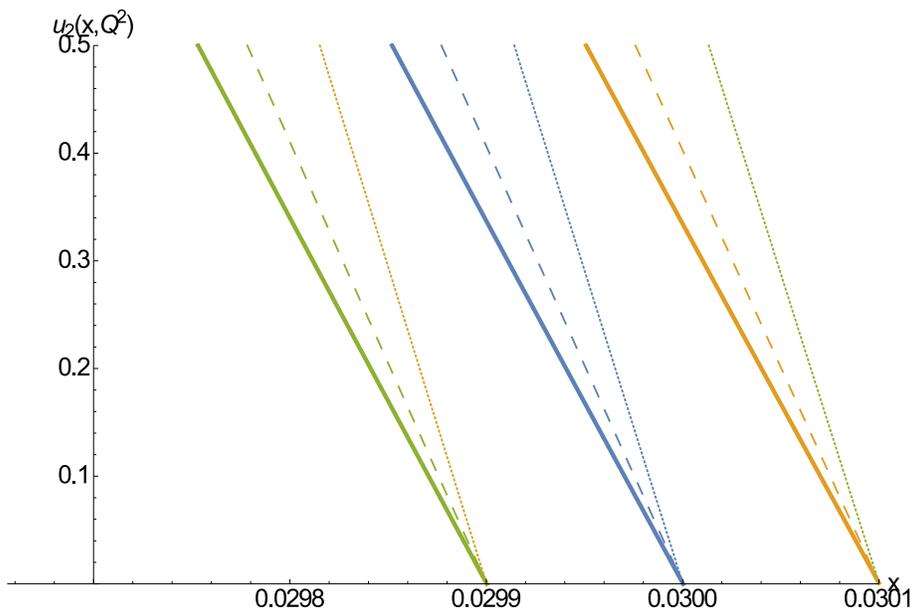}
\end{center}
  \caption{\label{figu2bessel} $u_2$ as a function of $x, Q$ (solid, dotted and dashed lines for $Q=1,2,3$ GeV) when started as a small value at a ``moderate'' $x$ (where processes such as higher\redd{-}twist can occur) and evolved to $x \rightarrow 0$.  The evolution is cut off when the linear approximation becomes untenable.
}
\end{figure}

For the Bessel parametrization, 
we start at a critical moderately high $x$ to fix the parameters characterizing the theory ($A_0,B_0$ for the Bessel function) \redd{in order } to reproduce $u_2/\epsilon \sim 10^{-2}$, in line with expectation from semi-perturbative higher\redd{-}twist processes.
We then go lower in $x$ of both projectile and target to analyze the behavior of $u_2,v_2$ as one gets closer to mid-rapidity.   As can be seen in Fig. \ref{figu2bessel} the instability indeed grows extremely rapidly from a broad value of Bjorken $x$.  In fact, very quickly it becomes so large that the linearization ansatz we used becomes inapplicable.

As a phenomenological fix to deal with this issue, we shall assume corrections beyond the leading order ``saturate'' $u_2$ (this time we exhausted the dimensions where instabilities can grow, so saturation is a plausible alternative).  An ad-hoc but consistent way to do this is replacing

\begin{equation}
\label{satu2}
u_2(x,Q) \rightarrow u_2^{max} \tanh\left( \frac{u_2(x,Q) }{u_2^{max}} \right).
\end{equation}

\redd{A}t low $u_2$, this recovers the linearized equations, but never goes above the parameter $u_2^{max}$.  Physically, this will be the scale where higher order corrections, assumed here to essentially cutoff $u_2$ growth, take over.  The price for this adjustment is that we lose \redd{the predictive power } of $u_2$ at a given $Q, x$ and centrality.

To convert these asymmetric distribution functions to something that can be related to experimental measurement, we would need to perform microscopic quark-and gluon scattering as well as use fragmentation functions.  This is a somewhat 
involved calculation, which, in the limit of low $x$, is thought to simplify with the  $k_T$ factorization ansatz \cite{kovbook}.   The physical picture, which we caution the reader to regard as a rough approximation in the regime here, is that of the lowest-order process, two gluons from two different hadrons fusing into a third gluon.   If the transverse structure of the two gluons is considered, and an angular dependence is included, the resulting gluon rapidity distribution \redd{would } look like

\[
\frac{dN}{dp_T dy} \sim  \frac{1}{p_T^2} \int  k dk d\theta  \bigg[ f_A(x_A,k,\theta) f_B(x_B,p_T-k,\pi+\phi-\theta)
\]

\begin{equation}
\label{ktformula}
\times \delta\big( x_A +x_B-e^{-y} \big)\bigg]\bigg|_{x_{A,B}=\frac{p_T e^{\pm y}}{\sqrt{s}}} ,
\end{equation}

and will point, preferentially, {\em perpendicularly} to the reaction plane by kinematics, producing the same harmonic dependence w.r.t. the reaction plane as that expected from hydrodynamics.
Motivated by Eq. \ref{ktformula}, we set $u_2^{max}$ at a value of $0.5$, where its contribution to hadron production becomes comparable to the unperturbed saturation value.  We note that the experimental $v_2$ is curiously quantitatively similar to this limit when averaged over centrality, although accounting for the proportionality w.r.t. eccentricity when dynamics is dominated by saturation would be problematic.

Together with quark-hadron duality, $k_T$ factorization can be used to provide a calculation for $v_2$ as it was for multiplicity \cite{kln}.  The shifting of momentum, due to fragmentation, from parton $p_T$ to hadron $p_{Th}$, necessary to provide the right limit for $dN/dp_T$, can be accomplished by updating Eq. \ref{ktformula} using

\begin{equation}
\label{fragment}
\frac{dN_h}{dp_{hT} dy_h} = \int_0^1 dz \int_{\Lambda_{QCD}}^{Q_s}   d p_T \delta\left( z- \frac{p_{Th}}{p_T} \right) z D\left( z,pT\right)\left. \frac{dN}{dp_T dy}\right|_{p_T} .
\end{equation}

\redd{W}e do this with a ``string-breaking'' Schwinger effect-inspired Gaussian fragmentation function, 
$ D(z,pT)\sim (1-z)e^{-z^2}$, and a pQCD-like function incorporating infrared divergences, $D(z)=(1-z)z^{-2}$.   Note that the integration limits are for the {\em parton} to have momentum larger than $\Lambda_{QCD}$ and smaller than $Q_s$.

Therefore, we warn the reader not to regard the following results as anything more than a qualitative estimate, and concentrate on the {\em scaling} of $v_2$ rather than the numerical value.

\begin{figure}
  \epsfig{width=0.5\textwidth,figure=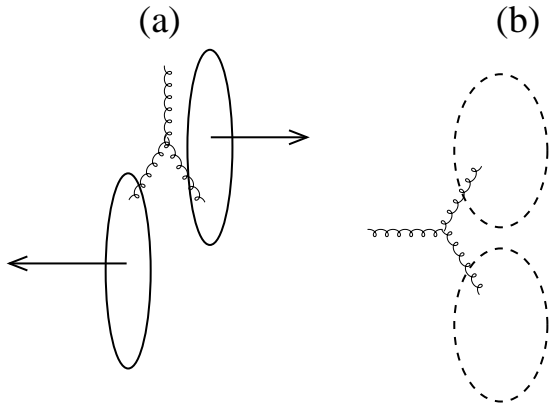}
  \caption{\label{ktfact}
    An illustration of how $k_T$ factorization converts asymmetric distributions of partons into asymmetric distributions of final gluons.  (a) and (b) show, respectively, the longitudinal and momentum-space transverse views of the collision
}
\end{figure}

\begin{figure}
  \epsfig{width=0.7\textwidth,figure=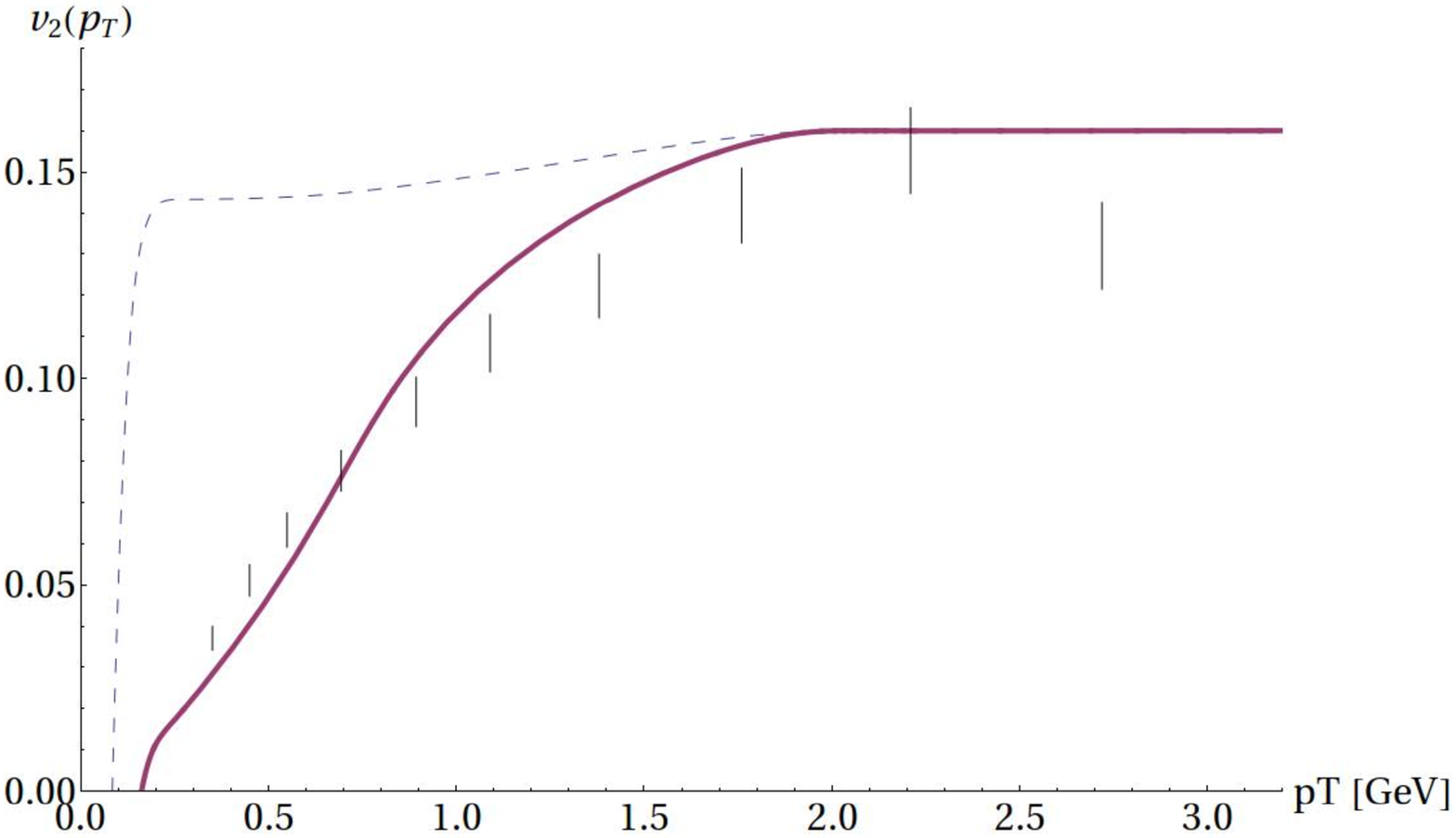}  
  \caption{\label{v2fig}  $v_2$ of partons (circles) and hadrons (lines) using $k_T$ factorization and fragmentation functions.  Dashed line gives a string-breaking ``Schwinger-based'' Gaussian fragmentation, solid line a QCD-like IR divergent fragmentation.   See text for details. A comparison with data \cite{cms1} is also shown
}
\end{figure}

The results are shown in Fig. \ref{v2fig}.   As can be seen, to capture the lower portion of the $v_2$ distribution, a correct treatment of fragmentation is essential. 

Parton-hadron duality badly misses the momentum dependence, with a nearly $p_T$ independent elliptic flow, although, in the spirit of structure functions, we have not included ``partons'' with momentum less than $\lqcd$.   A fragmentation with no divergence at low $z$ will still generally produce an unphysical discontinuity at $p_{Th}$, which can only be fixed with an infrared-divergent fragmentation function.
    The reader is  cautioned not to take this plot as anything except a rough qualitative estimate, especially since we have not included angular deflection during fragmentation, something known to be significant at lower momenta for particle fragmentations \cite{gtridge,esposito} and which would also be necessary to explain \cite{gtridge} the observed $v_2$ mass hierarchy \cite{meflork}.  
However, the rough qualitative agreement with the saturation value of $v_2$ is encouraging.

 Fig \ref{v2fig} shows a comparison with data, taken from \cite{cms1}, and confirms quantitative agreement.  Since we have no contact with geometry, the significance of this is merely that experimental $v_2$ is compatible with an appropriate $u_2^{max}$ for $k_T$ factorization, as well as an appropriate fragmentation function.


\section{Discussion and conclusions}

Unfortunately, while we certainly found instabilities in the GLR evolution,
our analysis has proven insufficient to make a successful contact with phenomenology at this stage.  The ansatz with realistic boundary conditions, section \ref{secbessel}, cannot have a proportional dependence on geometry because the instability grows too fast.  While saturating the growth gives an acceptable quantitative dependence, it is not clear how this saturation can depend on geometry.
The Polynomial ansatz of section \ref{secpol} fares better, since it has a quantitatively reasonable and qualitatively correct dependence on geometry, $Q$ and $x$.   It is, however, not clear how to implement physically motivated boundary conditions in this limit.

Thus, while two of our original motivations seem plausible, namely

\begin{itemize}
\item $u_n(Q)$ is practically independent on $x$.  Taking $k_T$ factorization seriously and assuming limiting fragmentation this means $v_2(p_T)$ has no explicit dependence on $y$ and $\sqrt{s}$, as in data.
\item The residual dependence of $v_n(p_T)$ is on $\lambda$ (see for example \cite{kln}), which is fit to the transverse multiplicity density $(1/S)(dN/dy)$ as in data. 
\end{itemize}

a phenomenological study of the viability of this model is left to further work. In particular, it remains to be seen whether realistic ``soft'' fragmentation breaks the scaling.   Such fragmentation is necessary to account for mass hierarchy \cite{gtridge}, commonly considered strong evidence for hydrodynamical behavior \cite{meflork}, but, because it is relevant for dynamics of soft gluons, little quantitative is known about it. 

From the theory side, our model is also somewhat simplified. To study instabilities quantitatively 
for extended systems, parton distribution functions will need to be promoted to include transverse structure \cite{generalized}.  The full dynamics governing this generalization will be a continuum non-linear partial differential equation, most likely not amenable to analytical solutions.
We note here that instabilities in 2+1 continuum partial differential equations typically generate an ``inverse Kolmogorov cascade'' \cite{cascade}, where low\redd{-}amplitude high\redd{-}frequency perturbations coalesce into low\redd{-}frequency high\redd{-}amplitude perturbations (3+1 dimensional instabilities typically behave the opposite way, low frequency/high amplitude $\rightarrow$ high frequency/low amplitude). \redd{Therefore, }   in an object extended in transverse area, many randomly pointing antennae, at the $x \rightarrow 1$ target/projectile, will tend to coalesce  into a single azimuthal instability at large $-\ln(1/x)$.   The exact form of this evolution, however, necessitates full control over the $\frac{\partial^2}{\partial Q \partial\phi}$ term in Eq. \ref{crossladder} and will therefore have to be left to a further work.

To compare our results to those of existing literature, we note that
while, superficially, the mechanism suggested here looks similar to other initial-state mechanisms for producing $v_2$ \cite{kovner,raju,bkb1,bkb2,bkb3}, the fact that $v_n$ is specifically generated by instabilities could be the key of avoiding the many randomly pointing antennae issue which makes these mechanisms untenable for systems with a large number of color domains, such as $v_n$ in $AA$ collisions.  It also ensures that $v_n$ is present in every cumulant of the particle distribution.

In case of a successful contact with phenomenology, 
an obvious way to experimentally test the ideas discussed in this work is via an $eA$ collider.  A prediction that can be made is that $eA$ collisions will give a non-zero $v_n$, perhaps comparable to the one given in hadronic collisions.   Of course, for such low multiplicity systems some care needs to be taken in defining $v_n$ in such a way that it does not include trivial non-flow effects such as energy momentum conservation and jet fragmentation.   A multiplicity of $\order{5-10}$ particles, together with a wide rapidity coverage, is required.   In this limit a ``ridge'' would show up as it does in $pp,pA$ collisions, as a multiparticle correlation elongated in rapidity both from the recoiling $e$ and between each of the few hadronic constituents.    If such a thing is found, it would form evidence that $G(...)$ does indeed have an initial azimuthal dependence.

Independently from the viability of the alternative proposed here, which, we underline, should be regarded at this point as a speculative suggestion, the phenomenological puzzles highlighted in section \ref{phenointro} remain, especially if a tighter scan in energy, rapidity, and system size (more high rapidity data together with lower energy $pA, dA$ data) cements the scaling shown in Eq. \ref{scale1} and \ref{scale2}.

Absent a comprehensive explanation of them from hydrodynamics, alternative ideas for the origin of $v_2$ should be explored.   Existing experimental data\redd{, } however\redd{, } imposes rather impressive challenges \redd{on new ideas for them to be considered } viable even qualitatively.    For instance, elliptic flow must depend on geometry, something now established experimentally by measurements of elliptic flow over rapidity \cite{phobos} , by \redd{measurements } of elliptic-flow and spectator correlations \cite{spectator}, and by the applicability of event-shape engineering ($v_2$ is proportional to geometry within each {\em event}, and this is an {\em experimental fact} \cite{engineering}).  Previous non-hydrodynamic explanations for $v_2$ \cite{kochvisc} did not have such a straight-forward relationship between geometry Fourier components and $v_n$ (in fact, rapidity-$v_2$ correlations were proposed as a signature \cite{kochhydro} for the hydrodynamic origin of $v_2$).  Neither would a generic azimuthal instability of parton distribution functions, or, for that matter, one triggered by Sivers-type processes (as explained in detail at the end of section \ref{phenointro}).    In a similar vein, the generation of $v_n$ must monotonically increase with transverse multiplicity density, independently from system size, for all cumulants, as confirmed by experimental data \cite{cmspp}.  Models whose source are localized color domain walls \cite{kovner,raju,pajares,antennae} can, at most,  describe $v_2$ for smaller systems, and have to deal with experimental evidence that small and large systems behave in a surprisingly similar way.
The instability scenario described in this work encouragingly could pass these obstacles.   Provided the instability is seeded by nuclear gradients, $v_n$ should be sensitive to initial geometry (See Fig. \ref{impact}) and be in the same direction with rapidity as it monotonically increases together with parton density. \redd{Also, } provided the instability in an extended system behaves as an inverse cascade, its dependence on the total multiplicity is likewise monotonically increasing, with different nuclei in a system extended in transverse space aligning the same way as partonic densities increase for smaller rapidities and denser systems.     This means the instability scenario could qualitatively describe the dependence of $v_n$ on both geometry and system size and is therefore a candidate for further development and quantitative testing.

In conclusion, we have discussed the phenomenological scaling of azimuthal anisotropy coefficients in hadronic collisions, arguing that it suggests an initial state origin compatible with ``Bjorken scaling'' phenomenology.  We have also noted that saturation dynamics corresponds also to a regime where azimuthal instabilities, in the full (2+1) evolution equations, could acquire growing modes.  We have used the azimuthally asymmetric GLR equation as a laboratory to test these growing modes, and found intriguing hints that they are indeed possible and have some qualitative features required for modeling $v_n$.   This suggests that this model should be developed further in order to connect it with data at a quantitative level, something that we plan to do in a subsequent work.

GT acknowledges support from FAPESP proc. 2014/13120-7 and CNPq bolsa de
produtividade 301996/2014-8. GG is supported by CNPq under the grant 151922/2014-4.  We would like to thank Al Mueller and Yuri Kovchegov for useful discussions and explanations.

\end{document}